\documentclass[aps,prd,onecolumn,nofootinbib,superscriptaddress]{revtex4}
\usepackage[colorlinks=true,linkcolor=blue,urlcolor=blue,filecolor=black,citecolor=red,pdfstartview=FitV,pdftitle={},pdfsubject={},pdfkeywords={},pdfpagemode=None,bookmarksopen=true]{hyperref}
\usepackage{graphicx}
\usepackage{amsmath}
\usepackage{amsfonts}
\usepackage{tensor}
\usepackage{amssymb,ulem}
\usepackage{color}%
\usepackage{dcolumn}
\usepackage{braket}
\usepackage{eurosym}
\usepackage[usenames,dvipsnames,svgnames]{xcolor}

\newcommand{\RNum}[1]{\uppercase\expandafter{\romannumeral #1\relax}}

\usepackage[title]{appendix}
\usepackage{wrapfig,boxedminipage,setspace,subfigure,epsfig}

\begin{document}
\baselineskip=0.5 cm

\title{Precessing and periodic timelike orbits and their potential applications in Einsteinian cubic gravity}

\author{Yong-Zhuang Li}
\email{liyongzhuang@just.edu.cn}
\affiliation{Research center for theoretical physics, School of Science, Jiangsu University of Science and Technology, Zhenjiang, China.}

\author{Xiao-Mei Kuang}
\email{xmeikuang@yzu.edu.cn (corresponding author)}
\affiliation{Center for Gravitation and Cosmology, College of Physical Science and Technology, Yangzhou University, Yangzhou, 225009, China}

\begin{abstract}
\baselineskip=0.4 cm
Einsteinian cubic gravity (ECG) is the most general theory up to cubic order in curvature, which have the same graviton spectrum as the Einstein theory. In this paper, we
investigate the geodesic motions of timelike particles around the four dimensional asymptotically flat black holes in ECG, and discuss their potential applications when  connecting them with recent  observational results. We first explore the effects of the cubic couplings on the marginally bound orbits (MBO), innermost stable circular orbits (ISCO) and on the periodic orbits around the Einsteinian cubic black hole. We find that comparing to Schwarzschild black hole in general relativity, the cubic coupling enhances the energy as well as the angular momentum for all the bound orbits of the particles.
Then, we derive the relativistic periastron precessions of the particles and give a preliminary bound on the cubic coupling employing the observational result of the S2 star' precession in SgrA*. Finally, after calculating the periodic orbits' configurations, we preliminarily evaluate the gravitational waveform radiated from several periodic orbits in one complete period of a test object which orbits a supermassive Einsteinian cubic black hole.
Our studies could be helpful for us to better understand the gravitational structure of the theory with high curvatures.
\end{abstract}

\maketitle

\tableofcontents
\newpage

\section{Introduction}

Recently, Einsteinian cubic gravity (ECG) has been proposed
as the unique cubic theory of gravity that shares
its graviton spectrum with Einstein gravity and has a
dimension-independent coupling constant \cite{Bueno:2016xff}. This theory has attracted lots of interest in the fields of
black hole and wormholes physics \cite{Bueno:2017sui,Bueno:2017qce,Mehdizadeh:2019qvc,Emond:2019crr,Burger:2019wkq,KordZangeneh:2020qeg,Burger:2020prd,Frassino:2020prd,Adair:2020prd}, cosmology \cite{Arciniega:2018fxj,Marciu:2020ysf,Quiros:2020uhr}, holography \cite{Bueno:2018xqc,Li:2019auk,Elena:2021jhep,Edelstein:2022xlb} and so on.
In particular,  the numerical black hole solution with asymptotic flatness in ECG theory was constructed in \cite{Hennigar:2016gkm,Bueno:2016lrh}, and the analytical approximation of the
black hole metric was given in \cite{Hennigar:2018hza} by employing the general
parameterization for spherically symmetric metrics proposed in \cite{Rezzolla:2014mua}.
The gravitational lensing effects of this black hole was studied
in \cite{Hennigar:2018hza,Poshteh:2018wqy}, which shew distinguishable features with that in Einstein theory. Moreover,  the quasinormal modes of scalar, electromagnetic and Dirac fields and their Hawking radiation of this  black hole in
ECG were calculated in \cite{Konoplya:2020jgt} and it was found that the cubic corrections would suppress both real and imaginary part of quasinormal
mode frequencies and the intensity of Hawking radiation.

As is known that black holes are natural laboratories to test gravity in the strong field regime. Recent astrophysical progresses including gravitational waves (GW) \cite{LIGOScientific:2016aoc,LIGOScientific:2018mvr,LIGOScientific:2020aai}, black hole shadows \cite{EventHorizonTelescope:2019dse,EventHorizonTelescope:2019ths,EventHorizonTelescope:2019pgp,
EventHorizonTelescope:2022xnr,EventHorizonTelescope:2022xqj}  and center of our Milky Way \cite{Genzel:2010zy} are strong evidences of the existence of black holes in our Universe. Theoretical and observational studies of the properties in the vicinity of black holes have been booming since they provide potential tools to test
general relativity (GR) and modified gravity theories. Among these, the geodesics around the black holes attract extensive interest, and the bound orbits of timelike particles are important to explore the motions of stars around SgrA* and GW radiation during the black hole merger, which are helpful for us to further understand the essence of gravity. One important type of bound orbit is the precessing orbit, through which the Mercury goes  and is one of the early evidences of GR \cite{Willbook}. The precessing of the stars around supermassive black hole in SgrA* have been explored in \cite{Iorio:2011zi,Grould:2017bsw,DeLaurentis:2018ahr,
GRAVITY:2019tuf,DES:2019ltu} and references therein. In particular, the measurement of the precessing orbit of S2 star \cite{GRAVITY:2020gka} has been fitting in different theories and  used to constrain the model parameters \cite{Hees:2017aal,DellaMonica:2021xcf,Yan:2022fkr,DellaMonica:2023dcw}. Thus, the precessing of timelike particle becomes one of potential tools to test alternative theories of gravity.

Another important type of bound orbit is the periodic orbit, which is special because generic  orbits around black hole can be treated as small deviations from periodic orbits encoding fundamental information about orbits around a central black hole \cite{Levin:2008mq}. In particular,
in the inspiral stage of black hole merge, the two initial black holes of extreme mass ratio approach to each
other due to the gravitational wave radiation. This sector can be approximated as a timelike test particle orbiting a supermassive black hole, and is known as extreme mass ratio inspiral (EMRI) system. In this scenario, the periodic orbits act as successive orbit transition states and play an important role in the study
of the gravitational wave radiation \cite{Glampedakis:2002ya}. Consequently, the authors of \cite{Levin:2008mq} proposed a  zoom-whirl classification for the periodic
orbits of the massive particles, which is characterized by three topological integers $(z, w, v)$ representing zoom, whirl and vertex behaviors of the orbit, respectively.
These periodic orbits with zoom-whirl pattern was previously found to maybe helpful for faster computation of adiabatic EMRI \cite{Grossman:2011im}, which got further progress  more recently in \cite{Tu:2023xab}. It is noted that  the zoom-whirl pattern of a periodic orbits has yet been observed, but it is common accepted that they could give more information about strong-field properties of the spacetime than the precessing orbits. Therefore,  the periodic orbits of timelike particles and their
taxonomy have been extensively studied in \cite{Tu:2023xab,Levin:2008ci,Levin:2009sk,Healy:2009zm,Misra:2010pu,Pugliese:2013xfa,Babar:2017gsg,Wei:2019zdf,Rana:2019bsn,Wang:2022tfo,Mummery:2022ana,
Habibina:2022ztd,Deng:2020yfm,Lin:2021noq,Lin:2023rmo,Bambhaniya:2020zno,Zhou:2020zys,Zhang:2022zox,Azreg-Ainou:2020bfl,Li:2023djs,Wu:2023wld,Lim:2024mkb} and references therein.

Thus, the aim of this paper is to study the timelike geodesics, precessing and periodic orbits, and connect them with remarkable observations to further understand the gravitational aspects of ECG. Our results show that the cubic corrections have significant influence on the bound orbits, especially the precessing and periodic motions. Thus,
we use the observational results of the S2 star' precession in SgrA* to give a preliminary bound on the cubic coupling.
In addition, we study the periodic orbits and preliminarily exam the
gravitational waveform radiated from the periodic orbits of a test object around a supermassive black hole in ECG theory.
It will be shown that the gravitational waveform indeed reflect the zoom-whirl behavior of the periodic orbits, which is also modified by the cubic couplings.

The paper is organized as follows. In section \ref{sec-background}, we will briefly review the ECG theory, its analytically approximate black hole solution and the timelike geodesic. In section \ref{sec-bound orbit}, by analyzing the effective potential, we study the effects of cubic corrections on the bound orbits of the timelike particles in ECG. In section \ref{sec-periodic orbits}, we first obtain a preliminary bound on the cubic coupling with the use of the observational result of the S2 star' precession in SgrA*. Then, we analyze the periodic orbits of the particle, based on which we examine the gravitational wave radiations  in one complete period of a test object which orbits a supermassive Einsteinian cubic black hole. Section \ref{sec-conclusion} contributes to our conclusion and discussion. In this paper, we shall use the units $G_N=c=1$ unless we especially restate.

\section{Background and timelike geodesic}\label{sec-background}
In this section,  we shall  briefly review the asymptotically flat, static and spherically symmetric  black holes, known as the  continued fraction solution to the metric of ECG obtained in \cite{Hennigar:2018hza}. The ECG theory  has the Lagrangian density \cite{Bueno:2016xff}
\begin{equation} \label{eq:Lagrangian density}
{\cal S} = \frac{1}{2 \kappa} \left(-2 \Lambda + R \right) + \beta_1 \chi_{4} + \kappa \left(\beta_2 \chi_{6} + \lambda {\cal P} \right),
\end{equation}
where $\kappa=8\pi G$, $R$ is the Ricci scalar, $\beta_1$ and $\beta_2$ are coefficients, $\chi_{4}$ and $\chi_{6}$ are four- and six-dimensional Euler densities which correspond to the usual Lovelock terms, and $\lambda$ is the coupling constant denoting the weight of  the cubic term
\begin{eqnarray}
{\cal P} = 12 \tensor{R}{_\mu ^\rho _\nu ^\sigma} \tensor{R}{_\rho ^\gamma _\sigma ^\delta}\tensor{R}{_\gamma ^\mu _\delta ^\nu} + \tensor{R}{_\mu _\nu ^\rho ^\sigma}\tensor{R}{_\rho_\sigma ^\gamma^\delta}\tensor{R}{_\gamma_\delta^\mu^\nu}
 - 12 R_{\mu\nu\rho\lambda}R^{\mu \rho}R^{\nu\sigma} + 8 R_\mu^\nu R_\nu ^\rho R_\rho ^\mu .
\end{eqnarray}
The field equation for the gravity under  a static, spherically symmetric metric
\begin{equation}\label{eq:ds2}
ds^2= -f(r)dt^2+\frac{dr^2}{f(r)}+r^2(d\theta^2+\sin^2\theta d\phi^2)
\end{equation}
reads~\cite{Hennigar:2016gkm}
\begin{eqnarray}\label{eq:fr}
-(f-1)r - \lambda \bigg[\frac{f'^3}{3} + \frac{1}{r} f'^2 - \frac{2}{r^2} f(f-1) f'- \frac{1}{r} f f'' (rf' - 2(f-1)) \bigg] = 2 M.
\end{eqnarray}
 The numerical solution of the metric function to the field equation was found in \cite{Hennigar:2016gkm,Bueno:2016lrh} of which the authors gave the mass and Hawking temperature of the black hole as
\begin{eqnarray}\label{eq:massANDtemp}
M &=& \frac{r_h^3}{12 \lambda^2} \left[r_h^6 + (2 \lambda - r_h^4) \sqrt{r_h^4 + 4 \lambda} \right] \, ,\nonumber\\
T &=&\frac{r_h}{8 \pi \lambda} \left[ \sqrt{r_h^4 + 4 \lambda} - r_h^2 \right] \, ,
\end{eqnarray}
with $r_h$ the radius of the event horizon.

Instead of the numerical black hole solution, in this paper we will focus on the analytic solution with the continued fraction method, in which the metric function is represented as \cite{Hennigar:2018hza}
\begin{equation}
\label{eq:f-analytical}
f(x) = x \left[1 - \varepsilon(1-x) + (b_0 - \varepsilon)(1-x)^2 + \tilde{B}(x)(1-x)^3 \right].
\end{equation}
Here $x$ is a new compact coordinate defined as  $x=1-r_h/r$ and  {$\varepsilon=2 M/r_h - 1$ is a small, positive quantity} determining the deviation of the radius of
the event horizon from the Schwarzschild radius. {The function}
\begin{equation}
\tilde{B}(x) = \cfrac{b_1}{1+\cfrac{b_2 x}{1+\cfrac{b_3 x}{1+\cdots}}}
\end{equation}
is a continued fraction of which  the coefficients can be determined from the field equations in term of $T$ and $M$ as
\begin{eqnarray}
&&b_0=0,~~ b_1 = 4 \pi r_h T + \frac{4 M}{r_h} - 3,~~b_2 = - \frac{r_h^3 a_2 + 16 \pi r_h^2 T + 6(M-r_h)}{4 \pi r_h^2 T + 4 M - 3 r_h} \, ,\\
&&b_3=\frac{1}{192 \pi  b_2 \lambda  r_h T \left(\pi  r_h T+\frac{1}{2}\right) \left(M+\pi  r_h^2 T-\frac{3
   r_h}{4}\right)}\Big[-128 \pi  \left(b_2+\frac{3}{2}\right)^2 \lambda  M^2 T+96 \pi  \left(b_2^2+b_2-1\right) \lambda
   M  T r_h
   \nonumber \\
&&\quad-448 \pi  \lambda  \left(\pi  \left(b_2^2+\frac{41
   b_2}{7}+\frac{93}{14}\right) M T-\frac{9 b_2}{28}-\frac{3}{4}\right)T r_h^2+240 \pi ^2 \left(b_2^2+\frac{31}{5} \left(b_2+1\right)\right) \lambda T^2 r_h^3
   \nonumber\\
&&\quad+ \left(-4 b_2 M-320 \pi ^3
   \left(b_2^2+7 b_2+16\right) \lambda  T^3-6 M\right)r_h^4+\left(3 b_2+6\right) r_h^5-4 \pi
   \left(b_2+3\right) T r_h^6\Big]\,, \cdots\cdots
\end{eqnarray}
and $b_2$ is related to a coefficient $a_2$ appearing in the near horizon expansion. It is addressed in \cite{Hennigar:2016gkm} that for small and moderate values of the coupling constant (saying $\lambda/M^4 \in [0, 5]$.), $a_2$ can be approximated as
\begin{eqnarray}
\label{eq:a2}
a_2^\star  \approx -\frac{1}{M^2} \frac{1 + 2.1347 ( \lambda / M^4) + 0.0109172 ( \lambda / M^4)^2}{4 + 15.5284 ( \lambda / M^4) + 8.03479 ( \lambda / M^4)^2},
\end{eqnarray}
and all higher order coefficients then can be determined by the field equations in terms of $T$, $M$, $r_h$ and $b_2$.

It is noted that the  representation \eqref{eq:f-analytical} is an approximation of
the numerical metric function in the whole space from the event horizon to infinity. This kind of representation was widely used to approximate numerical black hole
solutions, see for examples \cite{Konoplya:2019goy,Younsi:2016azx} and references therein, and it was found that the compact analytical form approximating the numerical metric has sufficient accuracy. With the above analytical forms at hands,  we can study the motions of particles around the black hole spacetime with the metric \eqref{eq:ds2} and \eqref{eq:f-analytical} which we will mention as Einsteinian cubic black hole for simplicity. For our purpose in this paper, it is enough to consider up to the third order expansion for the metric (i.e., we set $b_4=b_5=\cdots=0$.) and set the cubic coupling or weight constant to be positive.

Now we will derive the geodesic motion for a timelike particle moving on the equatorial plane of the spherically symmetric black hole. The Lagrangian of the timelike particle is \begin{eqnarray}\label{eqn:lag}
2\mathcal{L}=g_{\mu\nu}\dot{x}^\mu\dot{x}^\nu=-f(r)\dot{t}^2+\frac{\dot{r}^2}{f(r)}+r^2\dot{\phi}^2,
\end{eqnarray}
where the dot denotes the derivative with respect to the affine parameter $\tau$. Then it gives the equation
\begin{eqnarray}
p_t=\frac{\partial\mathcal{L} }{\partial \dot{t}}=-f(r)\dot{t}=-E,\\
p_\phi=\frac{\partial\mathcal{L} }{\partial \dot{\phi}}=r^2\dot{\phi}=-L,\\
p_r=\frac{\partial\mathcal{L} }{\partial \dot{r}}=\frac{1}{f(r)}\dot{r},
\end{eqnarray}
where $E$ and $L$ are the conserved energy and orbital angular momentum per unit mass of the particle, as the metric does not depend on the $t$ and $\phi$.
Considering that we can set the Hamiltonian $\mathcal{H}=p_\mu\dot{x}^\mu=-1/2$ for the timelike particle,
we obtain the radial equation of motion  as
\begin{equation}
\dot{r}^2=E^2-f(r)\left(1+\frac{L^2}{r^2}\right),
\end{equation}
from which we define the effective potential as
\begin{equation}
V_{\text{eff}}=E^2-\dot{r}^2=f(r)\left(1+\frac{L^2}{r^2}\right).
\end{equation}
Then as addressed in \cite{2004graa.book}, one can analyze various orbits of the particles around the black hole. In particular, $E=1$ is the upper limit for the bound orbits of the particles, since in this case we have $\dot{r}^2=E^2-V_{\text{eff}\mid_{r\to \infty}}=0$. Thus, the particles with $E>1$ can escape to infinity and bound orbits only exist when $E<1$.

\section{Bound orbits of timelike particle}\label{sec-bound orbit}

\begin{figure} [h]
{\centering
\includegraphics[width=8cm]{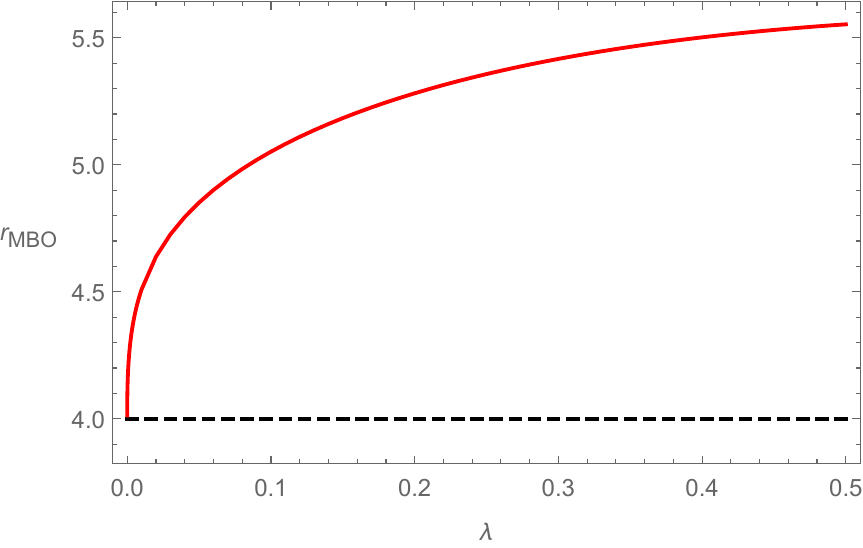}\hspace{5mm}
\includegraphics[width=8cm]{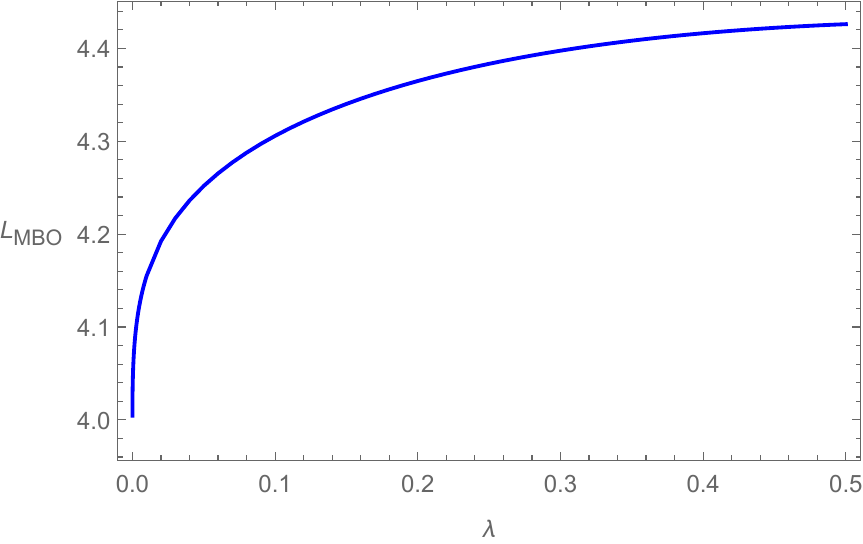}
   \caption{The radius and angular momentum for the marginal bound orbits around the Einsteinian cubic black hole.}   \label{fig:lambda-MBO}}
\end{figure}

In this section, we will study the properties of bound orbits of the massive particles in the cubic gravity, which exist between the MBO and the ISCO.
The MBO is a type of unstable circular orbit around the black hole, which has maximum energy $E=1$. It satisfies the conditions
\begin{eqnarray}
V_{\text{eff}}=1,~~~\frac{dV_{\text{eff}}}{dr}=0,
\end{eqnarray}
from which we can solve out the radial distance $r_{MBO}$ and the angular momentum $L_{MBO}$ of the particle around
the black hole with respect to the parameter $\lambda$ in the cubic gravity. The results are shown in FIG. \ref{fig:lambda-MBO}, from which we see that comparing to the Einstein gravity, both $r_{MBO}$ and $L_{MBO}$ become larger in the cubic gravity. It is noted that in our theoretical evaluations, all the physical quantities are rescaled by $M$ to be dimensionless, therefore, their features are independent of $M$ such that we set $M=1$ without loss of generality.
\begin{figure} [h]
{\centering
\includegraphics[width=5.5cm]{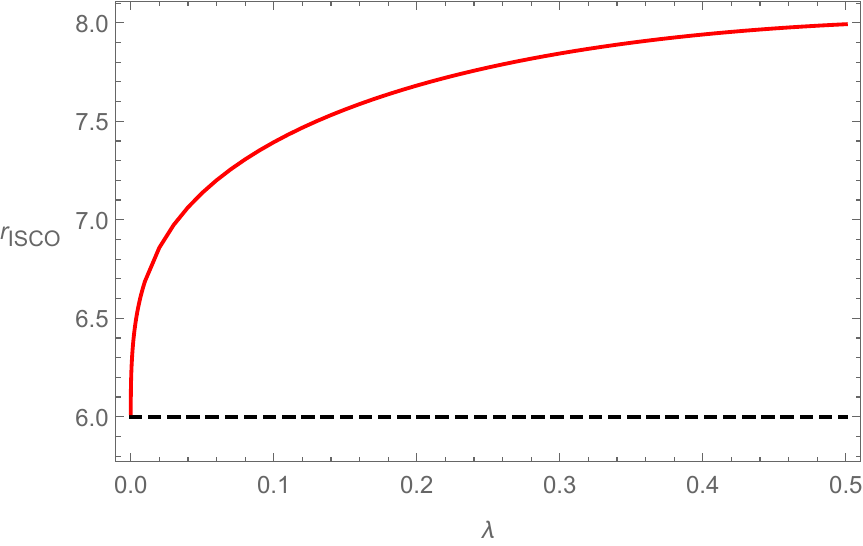}\hspace{1mm}
\includegraphics[width=5.5cm]{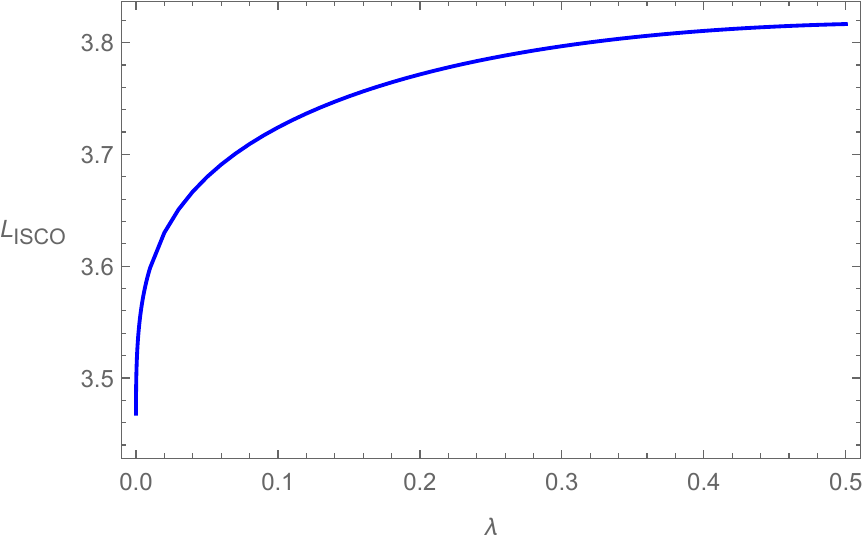}\hspace{1mm}
\includegraphics[width=5.5cm]{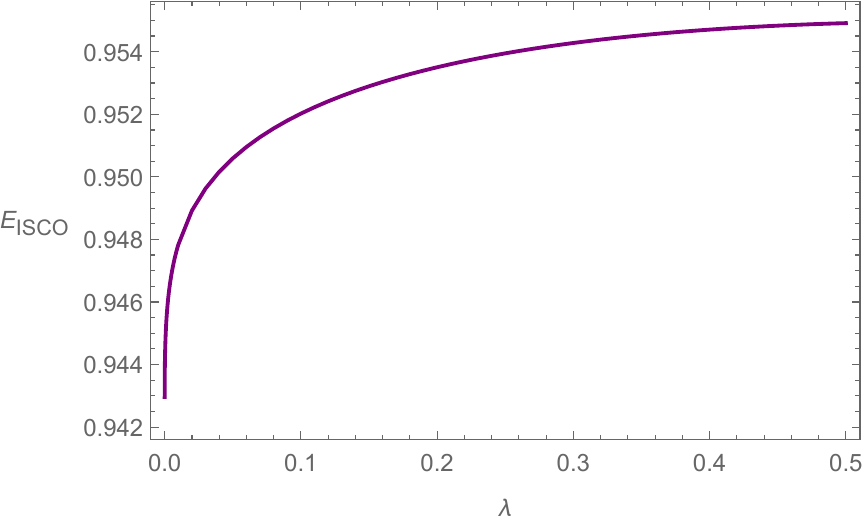}
   \caption{The radius, angular momentum and energy for the innermost stable circular orbits around the Einsteinian cubic black hole.}   \label{fig:lambda-ISCO}}
\end{figure}

On the other hand, ISCO is a type of marginal circular orbit around the black hole, which has the minimum allowed orbital radius. It is given by the conditions,
\begin{eqnarray}
V_{\text{eff}}=E^2,~~~\frac{dV_{\text{eff}}}{dr}=0,~~~\frac{d^2V_{\text{eff}}}{dr^2}=0.
\end{eqnarray}
Similarly, by solving the above equations group, we can determine the minimal radial distance $r_{ISCO}$, the angular momentum $L_{ISCO}$ and the energy  $E_{ISCO}$ of the particle as function of the parameter $\lambda$ in the cubic gravity.
It is obvious in FIG. \ref{fig:lambda-ISCO} that as coupling parameter $\lambda$ in the cubic gravity increases, $r_{ISCO}$, $L_{ISCO}$ and  $E_{MBO}$ all become larger. Our numerical results are consistent with the  approximation results analytically obtained in  \cite{Hennigar:2018hza}.
\begin{figure} [h]
{\centering
\includegraphics[width=10cm]{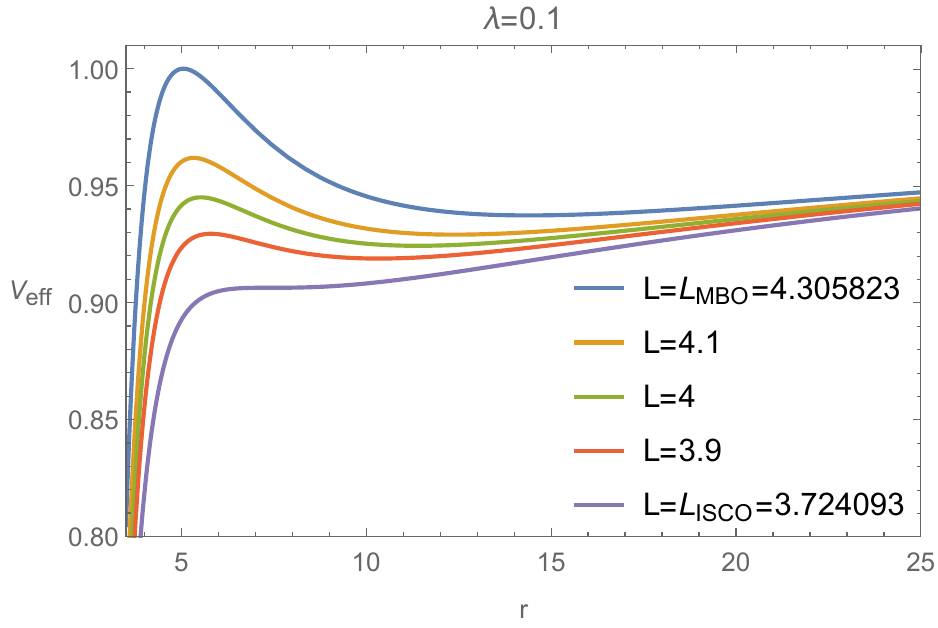}\hspace{5mm}
   \caption{The effective potential  of bound orbit for the Einsteinian cubic black hole with $\lambda=0.1$. The blue curve is for the MBO with $L=4.305823$ which has two extremal points, while the purple curve is for the ISCO with $L=3.724093$ which has one extremal point. }   \label{fig:r-Veff}}
\end{figure}
\begin{figure} [h]
{\centering
\includegraphics[width=10cm]{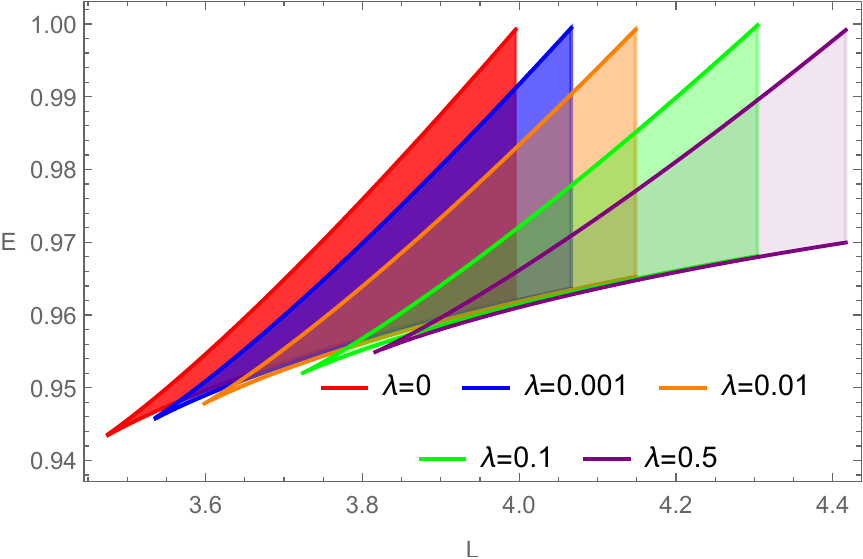}\hspace{5mm}
   \caption{The allowed $(L-E)$ regions for the bound orbits with selected coupling parameters in cubic gravity.}   \label{fig:L-E}}
\end{figure}
\begin{figure} [h]
{\centering
\includegraphics[width=12cm]{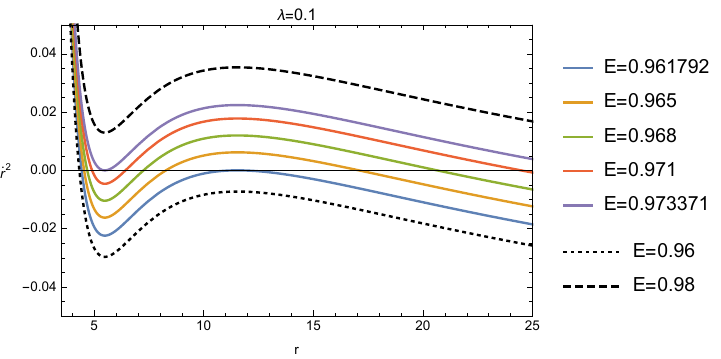}\hspace{5mm}
   \caption{The radial motion $\dot{r}^2$ as a function of $r$ with $\lambda=0.1$ and $L=\frac{L_{BMO}+L_{ISCO}}{2}=4.014958$ for selected energies. The bound orbits only exit when $\dot{r}^2=0$ has at least two roots. }   \label{fig:r-rdot}}
\end{figure}

Then, we can analyze the general properties of bound orbits locating between the MBO and ISCO around the Einsteinian cubic black hole, which we mainly extract from the effective potential and the radial motion of the timelike particle.  From the above studies, we know that for a fixed $\lambda$, the angular momentum for the bound orbits can only shift between the corresponding $L_{ISCO}$ and $L_{MBO}$, thus, we show the behavior of the effective potential for various bound orbits with $\lambda=0.1$ in FIG. \ref{fig:r-Veff}. The top curve corresponds to the effective potential of the MBOs with $L_{MBO}=4.305823$, and it has two extremal points, which will merge together as the angular momentum decreases to be $L_{ISCO}=3.724093$ in the bottom curve. In addition, the energy of bound orbits should also satisfy $E_{ISCO}<E<E_{MBO}$,  the allowed region of which also depends on the angular momentum of the particle. In FIG. \ref{fig:L-E}, we plot the allowed $(L-E)$ regions for the bound orbits. It shows that the allowed $(L-E)$ regions for the bound orbits will be modified by the cubic coupling comparing to the Einstein case.  Then, we move on to analyze the behavior of $\dot{r}^2$ for the particles with parameters in the allowed $(L-E)$ regions and beyond. The sample plots with $\lambda=0.1$ and $L=\frac{L_{BMO}+L_{ISCO}}{2}$ for different energies  are shown in FIG. \ref{fig:r-rdot}.  The bound orbits only exit in the regime $0.961792  \leq E \leq 0.973371$, corresponding that $\dot{r}^2=0$ has at least two roots. For the purple curve with $E=0.973371$, the bound orbits exit in the regime between the two roots. Then as we decrease the energy, there are three roots and  the bound orbits exist in the regime between the last two roots with $\dot{r}^2>0$, and the bound orbit converges to the ISCO when the energy reaches $E=0.961792$ corresponding to the blue curve. In addition, when the energy is larger (or smaller) than $E=0.961792~(E=0.973371)$,  $\dot{r}^2=0$ has no (or single) root which means no bound orbits, see for example the dashed (or dotted) curve.

\section{Precessing and periodic orbits}\label{sec-periodic orbits}
In this section, we will study special subclasses of bound orbits, the precessing and periodic orbits,  around the Einsteinian cubic black hole.
Due to the spherical symmetry of the background, the motions are completely determined by the $r-$ motion and $\phi-$ motion. For a bound orbit with two turning points $r_1$ and $r_2$, the timelike particle is reflected between the turning points. So, the apsidal angle $\Delta \phi$ passed by the particle in each period is
\begin{eqnarray}\label{eq:Dleta-phi}
\Delta \phi&=&\oint d\phi=2\int_{r_1}^{r_2}\frac{d\phi}{dr}dr.
\end{eqnarray}
Thus, according to \cite{Levin:2008mq}, the bound orbit can be described by a unique number $q$ as
\begin{equation}
q=\frac{\Delta\phi}{2\pi}-1.
\end{equation}
When $q$ is an irrational number, the timelike particle will move in precessing orbit and the precession per revolution is $\Delta\omega=\Delta \phi -2\pi$. Otherwise, it moves a periodic orbit with an rational number $q$, for which the timelike particles can return to its initial location after a finite time.  As addressed in \cite{Levin:2008mq}, a generic orbit can be treated as a perturbation of periodic orbits. Thus, it is significant to  study the periodic orbits which could give remarkable insight  for understanding the nature of any generic orbits and even gravitational radiation around the black hole.

Then based on $\Delta \phi$, we shall analyze the orbital precession and the preliminary bound on the cubic coupling with the use of S2 star's precession around SgrA*, and then figure out the periodic orbits to give the basic exploration of gravitational wave radiation from the periodic orbits.

\subsection{Precession and the preliminary bound on the cubic coupling}
Recently, GRAVITY published the Schwarzschild precession of the S2 star around Sgr A* by applying the spectroscopic
and astrometric measurements, and the ratio of the measured
Schwarzschild precession to the one predicted by general relativity is \cite{GRAVITY:2020gka}
\begin{eqnarray}\label{eq:fsp}
f_{sp}\equiv \frac{\Delta \omega_{s2}}{\Delta \omega_{GR}}=1.1\pm 0.19
\end{eqnarray}
where $\Delta \omega_{s2}$ and $\Delta \omega_{GR}$ are the observational precession and prediction from GR, respectively. It is obvious that the observation leaves some space for alternative gravity beyond GR, which could be used to constrain the model parameters.

Now we presuppose the central supermassive black hole in SgrA*
is the Einsteinian cubic black hole. The motion of S2 star is described by the trajectory \cite{2004graa.book}
\begin{eqnarray}\label{eq:r(Psi)}
r=\frac{a(1-e^2)}{1+e\cos \Psi}
\end{eqnarray}
reflected between the periastron, $r_p=a(1-e)$, and apastron, $r_a=a(1+e)$ as the turning points, where $e$ and $a$ are the eccentricity and the semi-major axis
of the orbit, and $\Psi$ is intersection angle between the semi-major axis and radial of the orbit.
Thus, the angle $\Delta\phi$ in \eqref{eq:Dleta-phi} can be rewritten as
\begin{eqnarray}\label{eq:Delta-phi}
\Delta \phi=2\int_{r_p}^{r_a}\frac{d\phi}{dr}dr=2\int_{0}^{\pi}\frac{d\phi}{d\Psi}d\Psi
\end{eqnarray}
where
\begin{eqnarray}\label{eq:dphidPsi}
\frac{d\phi}{d\Psi}=\frac{d\phi/dr}{d\Psi/dr}=
\frac{a e(1-e^2)L\sin\Psi}{r^2(1+e\cos\Psi)^2\sqrt{E^2-f(r)\left(1+\frac{L^2}{r^2}\right)}}.
\end{eqnarray}

\begin{figure} [h]
{\centering
\includegraphics[width=14cm]{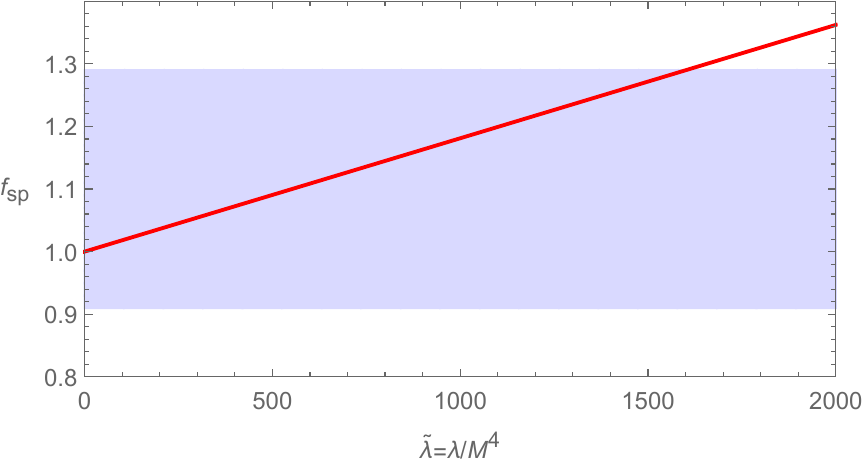}\hspace{5mm}
   \caption{The ratio of the measured precession by Einsteinian cubic gravity to the one by general relativity.}   \label{fig:lambda-fsp}}
\end{figure}

Considering that $\dot{r}\mid_{r=r_p}=0$ and $\dot{r}\mid_{r=r_a}=0$, we can solve $L$ and $E$ in terms of the turning points as
\begin{eqnarray}\label{eq:EL}
E^2=\frac{f(r_a)f(r_p)(r_p^2-r_a^2)}{r_p^2f(r_a)-r_a^2f(r_p)},~~~
L^2=\frac{r_a^2r_p^2(f(r_p)-f(r_a))}{r_p^2f(r_a)-r_a^2f(r_p)}.
\end{eqnarray}
Then substituting \eqref{eq:EL} into \eqref{eq:dphidPsi} and using the small $\lambda$ approximation of the metric function \cite{Hennigar:2018hza}
\begin{eqnarray}
f(r)=1-\frac{2M}{r}-\lambda\frac{1419(r/M)^2-8362r/M+10136}{12(65-61r/M)r^4},
\end{eqnarray}
we can evaluate \eqref{eq:dphidPsi} in the weak gravitational field as
\begin{eqnarray}\label{eq:dphidPsi2}
\frac{d\phi}{d\Psi}=1+\frac{(3+e\cos\Psi)M}{a(1-e^2)}+\tilde{\lambda}
\frac{473(3+e\cos\Psi)M^2}{488a^2(1-e^2)^2}+\mathcal{O}
\end{eqnarray}
where $\tilde{\lambda}=\lambda/M^4$ is the dimensionless coupling parameter. Subsequently, we put \eqref{eq:dphidPsi2} into \eqref{eq:Delta-phi} and obtain
\begin{equation}
\Delta\omega_{ECG}\simeq\Delta\phi-2\pi=\frac{6\phi M}{a(1-e^2)^2}+\frac{1419 \phi M^2 \tilde{\lambda}}{244 a^2(1-e^2)^2},
\end{equation}
in which when $\tilde{\lambda}\to 0$, we recover the result
$\Delta\omega_{GR}=\Delta\phi-2\pi=\frac{6\phi M}{a(1-e^2)^2}$ for Schwarzschild black hole in GR \cite{Weinberg:1972book}.
$f_{sp}$ in \eqref{eq:fsp} is then evaluated as
\begin{eqnarray}\label{eq:fsp2}
f_{sp}\equiv \frac{\Delta \omega_{ECG}}{\Delta \omega_{GR}}\simeq
1+\tilde{\lambda}\frac{473 M}{488 a(1-e^2)}.
\end{eqnarray}

Using $M_{SgrA}=4.3\times10^6 M_{\odot}$ with $M_{\odot}$ the mass of sun, $D=8.35\, kpc, a=125.058\,mas$ and $e=0.884649$ \cite{GRAVITY:2020gka}, we show the result of
$f_{sp}$ as a function of $\tilde{\lambda}$ (red curve) in FIG.\ref{fig:lambda-fsp} where the blue region denotes the range $1.1-0.19\leq f_{sp}\leq 1.1+0.19$. It implies that using the orbital precession of S2 star cannot rule out the Einsteinian cubic gravity and it can give a constraint $0\leq\tilde{\lambda}\leq1601.66$ which corresponds to $0<\lambda<5.47577\times10^{29} M_{\odot}^4$. It is noted that our constraint is more relax than $\lambda<4.57\times10^{22}M_{\odot}^4$ given in \cite{Hennigar:2018hza} with the use of Shapiro time delay experiment \cite{Will:2014kxa}.

\subsection{Seeking the periodic timelike orbits}
\begin{figure} [h]
{\centering
\includegraphics[width=8cm]{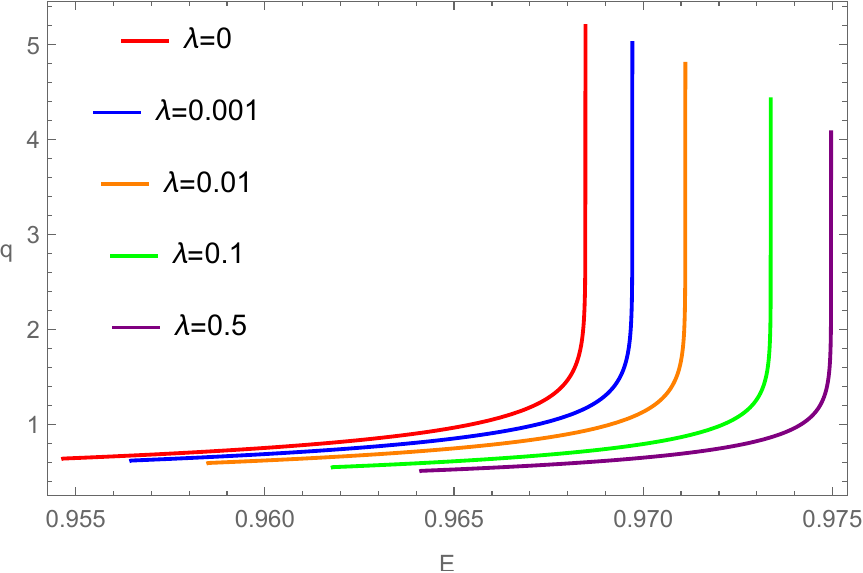}\hspace{5mm}
\includegraphics[width=8cm]{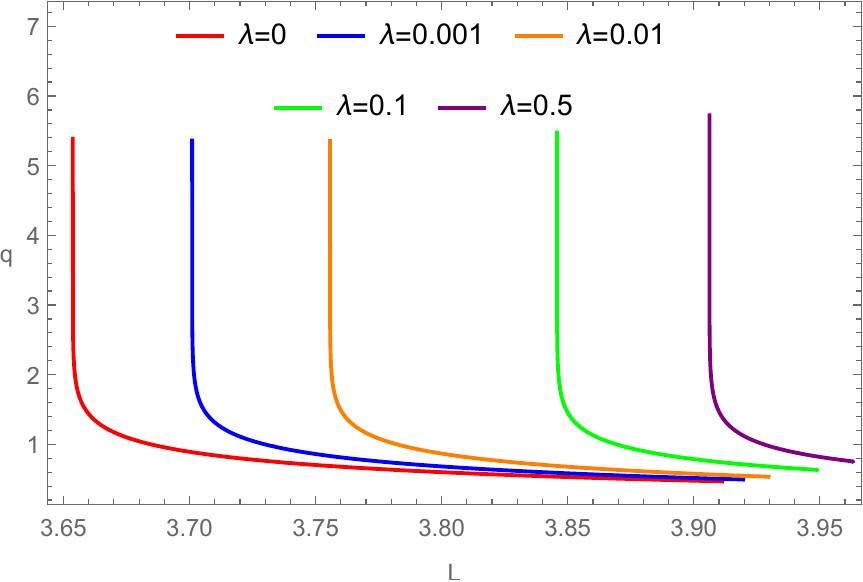}
   \caption{Left: the rational number $q$ as function of the energy for selected $\lambda$. We fix the angular momentum as $L=\frac{L_{BMO}+L_{ISCO}}{2}$. Right: the rational number $q$ as function of the angular momentum for selected $\lambda$. We fix the energy as $E=0.96$.}   \label{fig:q}}
\end{figure}

As we mentioned previously, the periodic orbits around the Einsteinian cubic black hole should be determined by $r-$motion and $\phi-$motion, the ratio between whose frequencies of oscillations  has to be a rational number. As suggested in the taxonomy of \cite{Levin:2008mq}, we define the ratio $q$ between the two frequencies in terms of  three integers (the zoom number $z$, the whirl number $w$ and the vertex number $v$  in the periodic orbits ) as
\begin{equation}
q=\frac{\Delta\phi}{2\pi}-1=w+\frac{v}{z},
\end{equation}
where $\Delta\phi$ is defined in \eqref{eq:Delta-phi}. So recalling the timelike geodesic equations of Einsteinian cubic black hole, we can calculate $q$ by
\begin{eqnarray}\label{eq:q}
q=\frac{\Delta\phi}{2\pi}-1=\frac{1}{\pi}\int_{r_1}^{r_2}\frac{d\phi}{dr}dr-1
=\frac{1}{\pi}\int_{r_1}^{r_2}\frac{L}{r^2\sqrt{E^2-f(r)\left(1+\frac{L^2}{r^2}\right)}}dr-1.
\end{eqnarray}
In FIG.\ref{fig:q}, we show the results of the rational number $q$ for periodic orbits as functions of the energy and angular momentum, respectively. From the left plot with fixed angular momentum  $L=\frac{L_{BMO}+L_{ISCO}}{2}$, for each selected cubic coupling, $q$ first increases smoothly as the energy increases, and then explodes as $E$ approaches its maximum values, similar as that occurs in the Schwarzschild black hole. For larger $\lambda$, $q$ is smaller in the allowed regime of energy. From the right plot with fixed $E=0.96$, for each selected cubic coupling, as the angular momentum decreases, $q$ also increases smoothly and then explodes as $L$ approaches its minimal values. However, $q$ of the bound orbits around Schwarzschild black hole can have smaller value in the allowed regime of $L$ than that around Einsteinian cubic black hole.
\begin{table*}[h!]
	\center
\begin{tabular}{|c|c|c|c|c|c|c|}\hline
 $\lambda$ & $E_{(1, 1, 0)}$ & $E_{(1, 2, 0)}$ & $E_{(2, 1, 1)}$ & $E_{(2, 2, 1)}$ & $E_{(3, 1, 2)}$ & $E_{(3, 2, 2)}$ \\ \hline
 0 & 0.965425 & 0.968383 & 0.968026 & 0.968434 & 0.968225 & 0.968438 \\ \hline
 0.001 & 0.967217 & 0.969669 & 0.96939 & 0.969706 & 0.969549 & 0.969709 \\ \hline
 0.01 & 0.969146 & 0.971084 & 0.97088 & 0.971109 & 0.970998 & 0.971111 \\ \hline
 0.1 & 0.972123 & 0.973358 & 0.973246 & 0.973369 & 0.973313 & 0.973370 \\ \hline\hline
\end{tabular}
\caption{The energy $E$ for the orbits with samples of $(z,w,v) $ for different cubic couplings. In each case, we fix $L=\frac{L_{BMO}+L_{ISCO}}{2}$.} \label{table:lambda-E}
\end{table*}

\begin{table*}[h!]
	\center
\begin{tabular}{|c|c|c|c|c|c|c|}\hline
 $\lambda$ & $L_{(1, 1, 0)}$ & $L_{(1, 2, 0)}$ & $L_{(2, 1, 1)}$ & $L_{(2, 2, 1)}$ & $L_{(3, 1, 2)}$ & $L_{(3, 2, 2)}$ \\ \hline
  0 & 3.683733 & 3.653557 & 3.657746 & 3.652853 & 3.655453 & 3.652788 \\\hline
 0.001 & 3.729857 & 3.701785 & 3.705602 & 3.701157 & 3.703533 & 3.701100 \\\hline
 0.01 & 3.782267 & 3.756387 & 3.759838 & 3.755831 & 3.757960 & 3.755781 \\\hline
 0.1 & 3.869314 & 3.846339 & 3.849378 & 3.845848 & 3.847724 & 3.845805 \\\hline\hline
\end{tabular}
\caption{The angular momentum $L$ for the orbits with samples of $(z,w,v) $ for different cubic couplings. Here we fix $E=0.96$.}\label{table:lambda-L}
\end{table*}

\begin{figure} [h]
{\centering
\includegraphics[width=12cm]{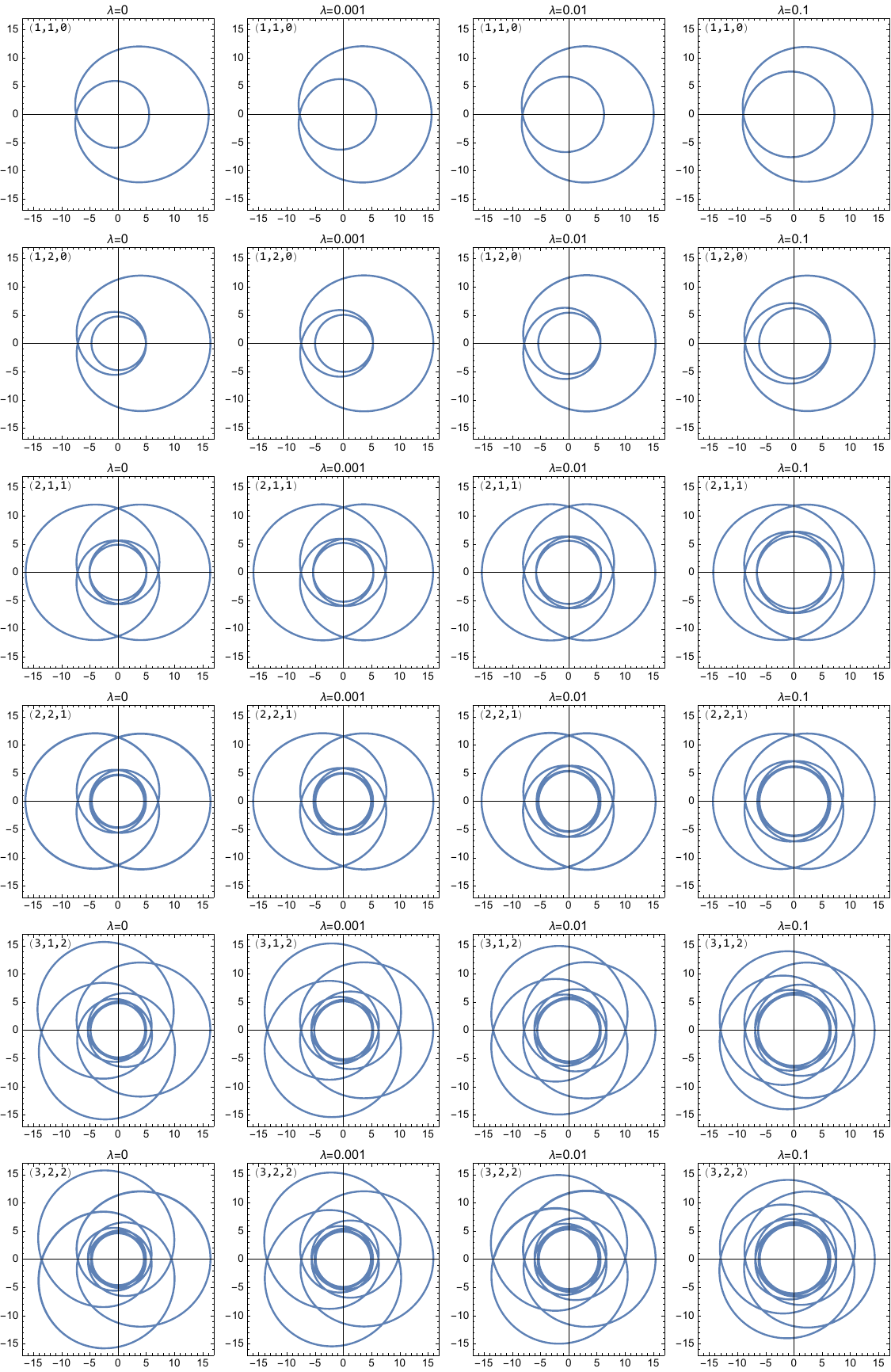}
   \caption{Periodic orbits with samples of $(z,w,v)$ for different cubic couplings. We fix the energy as $E=0.96$ and the corresponding $L$ is listed in Table \ref{table:lambda-L}.}   \label{fig:periodicOrbit}}
\end{figure}

Before giving that how the periodic orbits $(z,w,v)$ are affected by the cubic coupling, we list their corresponding energies $E_{(z,w,v)}$ with fixed angular momentum $L=\frac{L_{BMO}+L_{ISCO}}{2}$, and their corresponding angular momenta $L_{(z,w,v)}$ with fixed $E=0.96$ in TABLE \ref{table:lambda-E} and \ref{table:lambda-L}, respectively. The tables indicate that for each periodic orbit, both $E_{(z,w,v)}$ and $L_{(z,w,v)}$ increase for larger $\lambda$, implying that the particle
orbiting a Einsteinian cubic black hole always has higher energy or angular momentum with the other fixed than that in GR.
Then, we are ready to figure out the periodic orbits with $(z,w,v)$ in the polar coordinates $(r,\phi)$. As examples, we show the results with fixed $E=0.96$ in FIG. \ref{fig:periodicOrbit} of which the corresponding $L$ is present in Table \ref{table:lambda-L}. Comparing the orbits with the same $(z,w,v)$  in each row, we find that as the cubic coupling increases, the outmost trajectory  becomes smaller. It is obvious that $z$ describes the number of the leaf pattern of the orbit. By comparing the orbits with different $(z,w,v)$ in each column, we see that as the increase of $z$, the leaf pattern grows such that the orbit becomes more complicate.

\subsection{Gravitational wave radiation from periodic orbits}

We move on to study the gravitational radiation emitted by the periodic orbits of a test particle orbiting an Einsteinian cubic black hole. It means that we actually mimic an EMRI system, in which the smaller object has a mass {extremely} smaller than the central black hole.
Therefore, one can treat this small object as a perturbation to the spacetime of the Einsteinian cubic black hole. Then due to the gravitational radiation, the energy and angular momentum of the object will change over the periodic motions but small enough to consider the adiabatic approximation.  Thus, as addressed in \cite{Tu:2023xab}, we can trace the previous
periodic orbits of the object and calculate the corresponding gravitational wave radiation in this scenario.

\begin{figure} [h]
{\centering
\includegraphics[width=16cm]{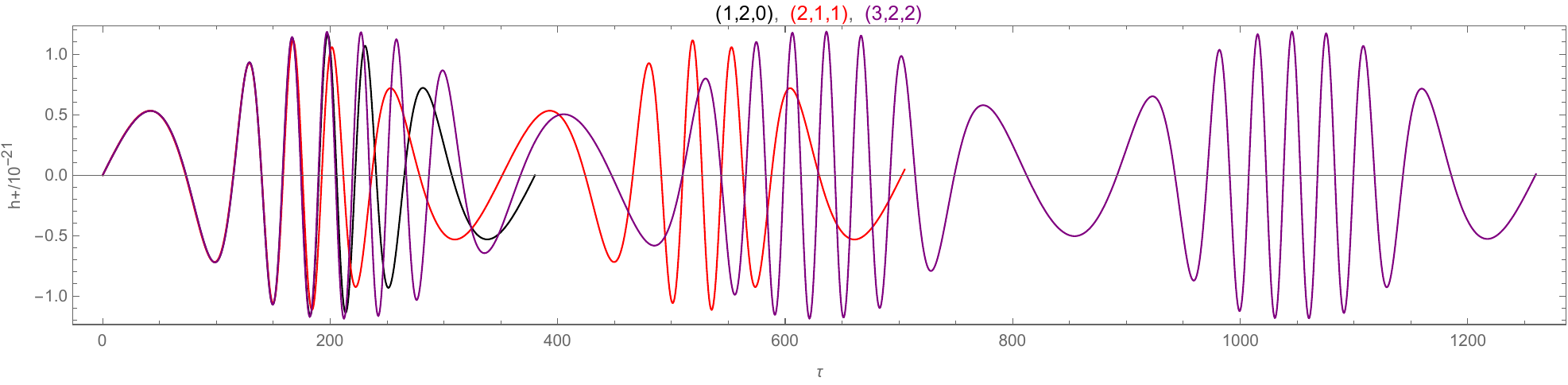}\hspace{2cm}
\includegraphics[width=16cm]{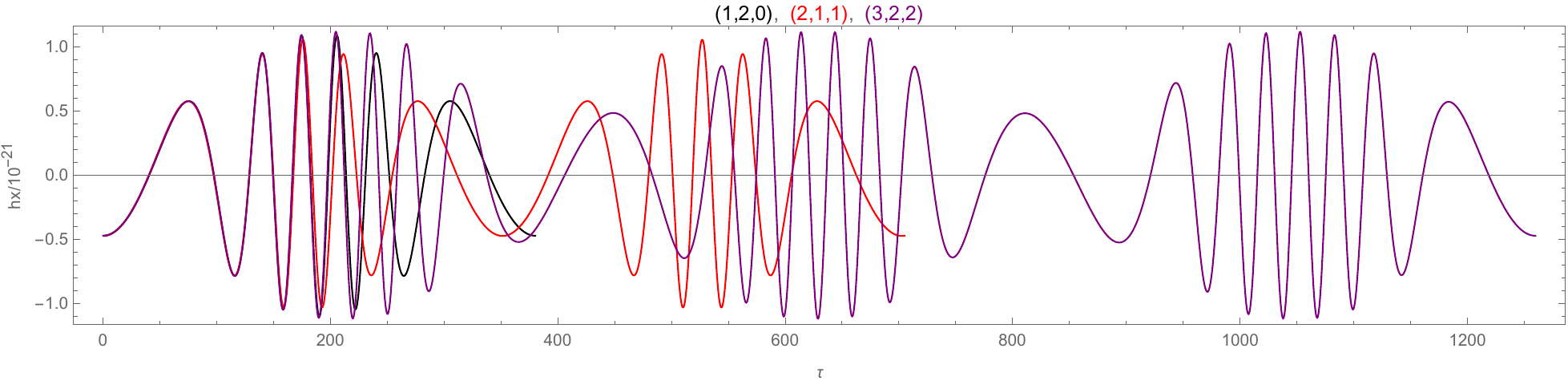}
   \caption{Gravitational waves for various periodic orbits. We fix $\lambda=0.1$ and the energy as $E=0.96$.}   \label{fig:GW1}}
\end{figure}

\begin{figure} [h]
{\centering
\includegraphics[width=16cm]{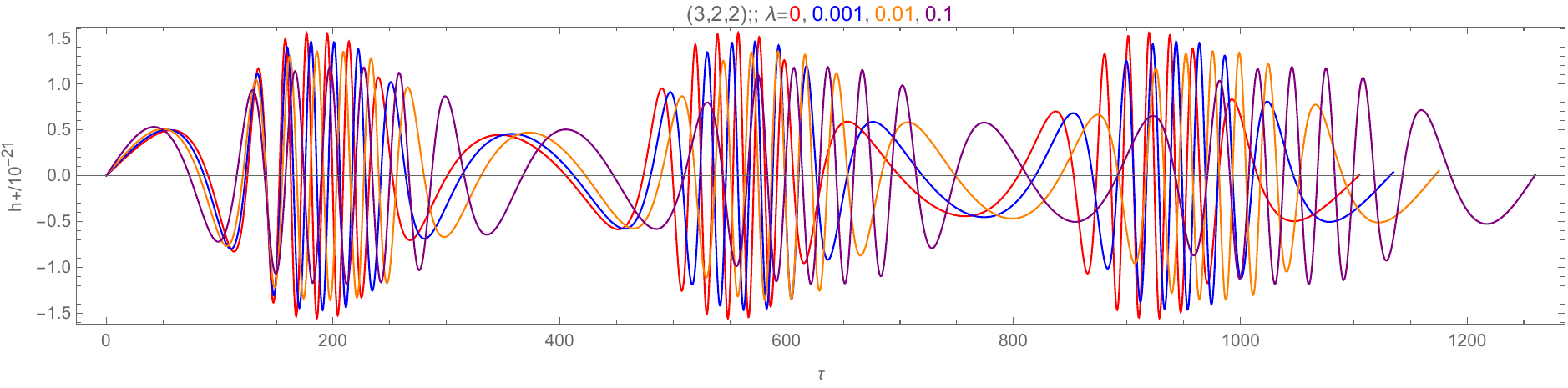}\hspace{2cm}
\includegraphics[width=16cm]{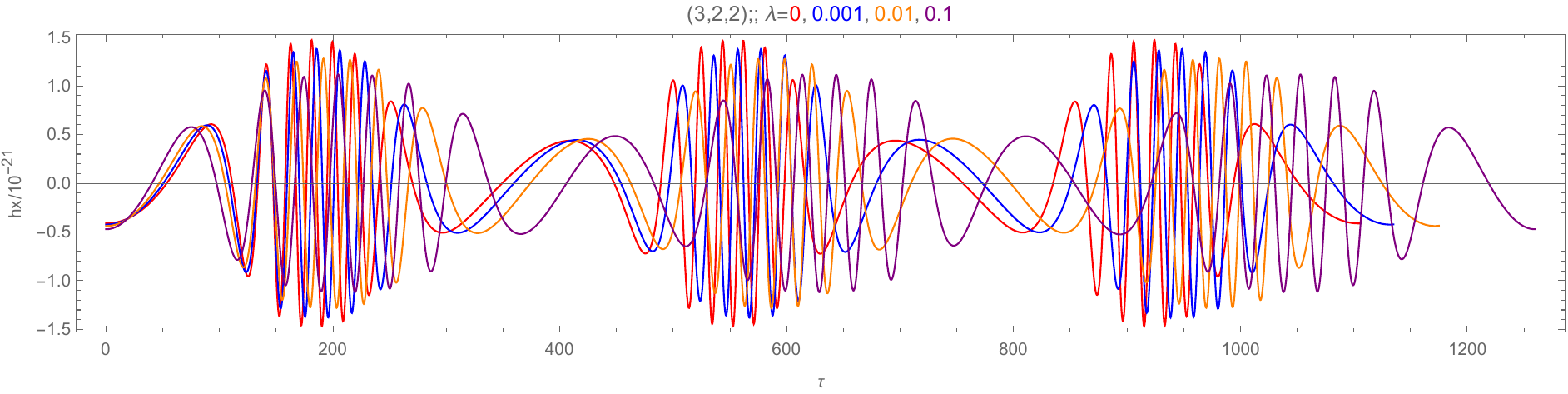}
   \caption{Gravitational waves for periodic orbit $(3,2,2)$ with different cubic coupling. We fix the energy as $E=0.96$.}   \label{fig:GW2}}
\end{figure}

To this end, we adopt the proposal in \cite{Tu:2023xab} and employ the kludge waveform developed in \cite{Babak:2006uv}. Then the gravitational waves emitted from the above periodic orbits in Einsteinian cubic black hole are calculated up to the quadratic order as \cite{Maselli:2021men}
\begin{equation}
h_{ij} = \frac{4 \eta M}{D_{\rm L}} \left(v_i v_j - \frac{m}{r} n_i n_j\right). \label{eq:quadratic}
\end{equation}
Here, $M$ and $m$ are the masses of the Einsteinian cubic black hole and the test particle, respectively. $D_{\rm L}$ denotes the luminosity distance of the EMRI system. $\eta=M m/(M+m)^2$ is known as the symmetric mass ratio. $v_i$ is the spatial velocity of the test particle. And $n_i$ is the radial unit vector of the motion of the test particle.
Then, we move on to project the above GW onto the detector-adapted coordinate system, in which the GW polarizations, $h_+$ and $h_\times$, take the forms  \cite{Maselli:2021men}
\begin{eqnarray}
h_+ &=& - \frac{2 \eta}{D_{\rm L}} \frac{M^2}{r} (1+\cos^2\iota) \cos(2\phi+2 \zeta), \\
h_\times &=& - \frac{4 \eta}{D_{\rm L}} \frac{M^2}{r} \cos\iota \sin(2\phi+2 \zeta).
\end{eqnarray}
Here, $\iota$ denotes the inclination angle between the EMRI's orbital angular momentum and the line of sight, and $\zeta$ denotes the latitudinal angle.

To analyze the gravitational waveform of different periodic orbits and how the cubic coupling affects the waveform, we  consider an EMRI system that consists of a supermassive Einsteinian cubic black hole with mass $M = 10^7 M_{\odot}$ and a small component with mass $m=10 M_{\odot}$. We set the luminosity distance $D_{\rm L} = 200\;{\rm Mpc}$, and  the inclination angle $\iota$ and the latitudinal angle $\zeta$  to be $\iota=\pi/4$ and $\zeta=\pi/4$.
Moreover, since we consider the adiabatic approximation and ignore the backreaction of the gravitational radiation to the periodic orbits, so we will only consider one complete period of the orbit in the  calculation to guarantee the accuracy. In fact, our calculations based on \eqref{eq:quadratic} cannot give us accurate waveforms for gravitational radiation because it excludes the contributions of multipoles higher than the quadratic order, but this is enough for us to capture the basic gravitational signals emitted from periodic orbits in ECG theory.
The $h_+$ and $h_{\times}$ of gravitational wave polarization from samples of periodic orbits with $E=0.96$ {and} $\lambda=0.1$ are shown in FIG. \ref{fig:GW1}. We see that the gravitational waveforms have corresponding zoom and whirl phases in one complete period, and give the zoom-whirl behaviors
of the periodic orbits. Then, we fix the periodic orbit $(3,2,2)$ and show the gravitational waveform for different cubic couplings in FIG. \ref{fig:GW2}. It is obvious that the cubic coupling affects the phase of gravitational waveform, and larger $\lambda$ suppresses the amplitudes of gravitational wave signals.

These preliminary studies imply that the gravitational wave radiated from the periodic orbit indeed
give the key information of whirl and zoom phases, so that it could be a potential tool to testify the properties
of the periodic orbits and further distinguish the Einsteinian
cubic gravity and GR.

\section{Conclusion and discussion}\label{sec-conclusion}
In this paper, we studied the geodesic motions of timelike particles around the four dimensional asymptotically flat black holes in Einsteinian cubic gravity, and discussed their potential observational applications. Our results indicate that the cubic coupling has a significant impact on the geodesic motions of timelike particles comparing to the Schwarzschild black hole. Meanwhile, the observational data of the S2 star in SgrA* cannot disprove the Einsteinian cubic gravity.

Taking the advantage of continued fraction approximations, we firstly investigated the general properties of bound orbits of the massive particles in the cubic gravity, which exist between the MBOs and the ISCOs. As expected, with an increasing positive coupling parameter in the cubic gravity, the energy, the angular momentum as well as the radius of MBOs and ISCOs all become larger for the bound orbits of the particles. The allowed $(L-E)$ regions for the bound orbits are then modified by the increasing coupling constant.

Then we studied two special subclasses of bound orbits, the precessing and periodic orbits. The former is a potential tool to constrain the model parameters of alternative theories of gravity while the latter encodes fundamental information about orbits around a black hole. Utilizing the observation data \cite{GRAVITY:2020gka} we relaxed the constraint of coupling parameter $\lambda$ to $0<\lambda<5.47577\times10^{29}M^{4}_{\odot}$, suggesting that the ECG can not be entirely excluded. The periodic orbits were investigated following the taxonomy \cite{Levin:2008mq}, and the results indicate that for fixed angular momentum $L$ the rational number has the similar behavior with the Schwarzschild black hole but will have a larger regime. Especially, for each periodic orbit the particle orbiting a static black hole in ECG always has the higher energy or angular momentum with the other fixed than that in GR, which is consistent with the previous section.

Finally, by mimicking an EMRI system we studied the gravitational radiation emitted by the periodic orbits of a test particle orbiting a supermassive Einsteinian cubic black hole. For the parameters we specified, we conclude that larger cubic coupling suppresses the amplitudes of gravitational wave signals, implying that the gravitational wave radiated from the periodic orbit indeed is significant in distinguishing the ECG and GR.

The foregoing preliminarily investigation have adopted a few approximations and assumptions for easing our analysis but will not change the main conclusions. It will be natural to discard these assumptions for more details about the geodesic motions of the timelike particles. Meanwhile, the current study could provide some insights for the future directions on the Einstein rings, the photon ring holography and so on in such ECG theory.

\begin{acknowledgments}
This work is partly supported by Natural Science Foundation of China under Grants No.12375054, Natural Science Foundation of Jiangsu Province under Grant No.BK20211601.
\end{acknowledgments}


\bibliography{ref}
\bibliographystyle{apsrev}

\end{document}